\documentclass[11pt]{llncs} 
\usepackage{amsmath}
\usepackage{subfigure}
\usepackage{graphicx}
\usepackage{alltt}
\usepackage{url}
\usepackage{xspace}
\usepackage{epsfig}

\usepackage{multirow}
\usepackage{color}

\usepackage[boxed,lined,longend,algo2e]{algorithm2e}
\usepackage[small,bf]{caption}
\usepackage{subsections}

\newcommand{\myurl}[1]{{\url{#1}}} 

\begin{document}

\title{Detecting Anomalous User Behavior Using an Extended Isolation Forest Algorithm: An Enterprise Case Study}

\author{Li Sun$^1$, Steven Versteeg$^2$, Serdar Bozta\c{s}$^1$， and Asha Rao$^1$}
\institute{$^1$ School of Mathematical and Geospatial Sciences, RMIT University,GPO Box 2476V, Melbourne, VIC 3001, Australia\\
$^2$ CA Labs, Melbourne, Australia}

\date{25 August 2016}
\maketitle


\begin{abstract}
Anomalous user behavior detection is the core component of many information security systems, such as intrusion detection, insider threat detection and authentication systems. Anomalous behavior will raise an alarm to the system administrator and can be further combined with other information to determine whether it constitutes an unauthorised or malicious use of a resource. This paper presents an anomalous user behaviour detection framework that applies an extended version of Isolation Forest algorithm.  Our method is fast and scalable and does not require example anomalies in the training data set. We apply our method to an enterprise dataset. The experimental results show that the system is able to isolate anomalous instances from the baseline user model using a single feature or combined features.
\end{abstract}

\section{Introduction}
\label{sec:intro} 

Detecting malicious activities is an important part of an
information security strategy. Attacks may come from intruders or
malicious insiders. While not every anomaly is an attack, nearly
every attack includes anomalous user behaviour. Effective detection
on anomalous user behaviour is therefore an important part of
detecting attacks.

Many existing machine learning methods for anomaly detection require examples of
both acceptable behaviour and malicious behaviour to train the system.
Finding a sufficient number of example attacks can be a challenge,
as these events by nature are relatively rare.

We propose a new method for detecting anomalies in user behaviour
based on the isolation forest anomaly detection algorithm.
Our method does not require any example anomalies in the training set.
We apply our method to an enterprise dataset of staff
accessing the payroll system in a large enterprise organisation.

In this paper, an effective anomalous user behavior detection system has been developed. The main contributions of this paper are towards 
\begin{itemize}
		\item We apply an extended version of Isolation Forest algorithm to isolate anomalous user behavior. We extend the Isolation Forest algorithm to support categorical data dimensions.
		
		\item We evaluate different features and combine less informative features together to achieve high detection effectiveness.

    \item We evaluate the anomalous user behavior detection system on a large number of log files that belongs to a real enterprise.
		
\end{itemize}
The remainder of the paper is organized as follows. Section~\ref{sec:related} provides research background on anomaly detection and discusses related work. Section~\ref{sec:methodology} gives the implementation of our system. Section~\ref{sec:experiment} describes the experimental setting and reports evaluation results. Section~\ref{sec:conclusion} discusses future directions and concludes the paper.

\section{Related Work}
\label{sec:related}
Anomaly user behavior detection techniques are designed to identify rare user behavior that are outliers to the normal behavior or behaviors that changed or carried by other users. These techniques can be categorized into three modes, supervised anomaly detection, semi-supervised anomaly detection and unsupervised anomaly detection. 

The supervised anomaly detection requires a set of labeled training data, normal or anomalous, to educate the detection software and create a predictive model of normal versus anomalous behavior. Once a new action is coming, the predictive model is applied to classify the new action. Examples of such approaches include support vector machines \cite{cb98:svm} and Bayesian networks \cite{tsk:dm06}. The problem with the supervised approach is that the system can only detect anomalies that conform with its model due to the lack of anomaly instances. A truly novel anomaly event may escape detection, as the training data contains no archetype for it. Additionally, huge effort required in labeling training data and human error in labeling will influence all future analysis.

Semi-supervised anomaly detection has the advantage of only requiring labeling for normal data within the training set. This works well for scenarios where anomaly types are difficult to predict, but `normal' is easy to define. In this mode, an anomaly is anything which lies outside of the normal labeled data. Semi-Supervised has the benefit of reducing the level of manual effort required to label the training data, as anomalies need not be tagged. This mode also has the benefit of being able to detect novel events, as again, anything not-normal is anomalous. Semi-supervised has difficulties when presented with noisy data sets. As noise may be distributed across the sample space, much of it may fall outside of the regions defined as normal, and hence this data will be declared anomalous and a high false-positive rate will result.

Unsupervised anomaly detection approach has the significant advantage of not requiring labeled training data. It operates on the assumption that normal data instances are the most common within the data set. Anomalies are rare. There may be other determining factors, such as clustering: normal data will congregate in large, dense clusters and anomalies occupy isolated positions within the data set. Unsupervised mode can be better at detecting novel events and can also recalibrate its definition of normal, should the normal-state within the input data evolve over time. Past work has also explored unsupervised learning for anomaly user behavior detection~\cite{lmhm05:ieee,eapps02:DM}. Cook \cite{ch07:IAW}, Eberle \cite{eh07:DMIN} and Yan \cite{yh02:ICDM} represent threat and non-threat data as a graph and apply unsupervised learning to detect anomalies. Our work focus on unsupervised approach. We represent user behaviour by combining features together and apply Isolation Forest algorithm \cite{liu2008isolation}, to build an automatic anomaly user detection system.

\section{Methodology}
\label{sec:methodology}
\begin{figure}[ht]
    \centering
    \includegraphics[height=3.55cm]{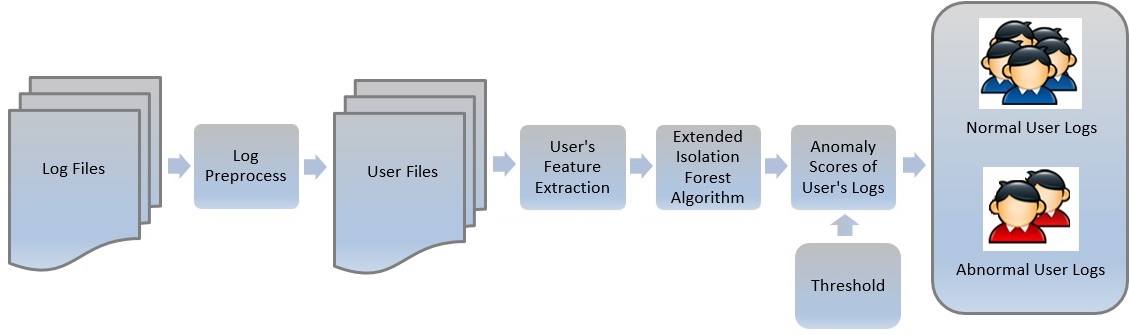}
    \caption{The proposed framework}
    \label{fig:frame}
\end{figure}

In this paper, we develop an automatic anomalous user detection system using an extended version of the Isolation Forest algorithm. The proposed framework is shown in Figure~\ref{fig:frame}. The parser pre-processes all log files and stores logs by users. For each user, the system extracts the features set and builds a baseline user model by creating a collection of extended Isolation Forest tree. When a new user record is coming, it is mapped into each of these Isolation Forest trees and an anomaly score is calculated. If its anomaly score is under the threshold, it is assumed normal, otherwise, an anomalous behavior is recorded and this user is flagged as anomalous.
 
\subsection{Isolation Forest}
\label{if}

\begin{figure}
\subfigure[]{\includegraphics[width=0.45\textwidth]{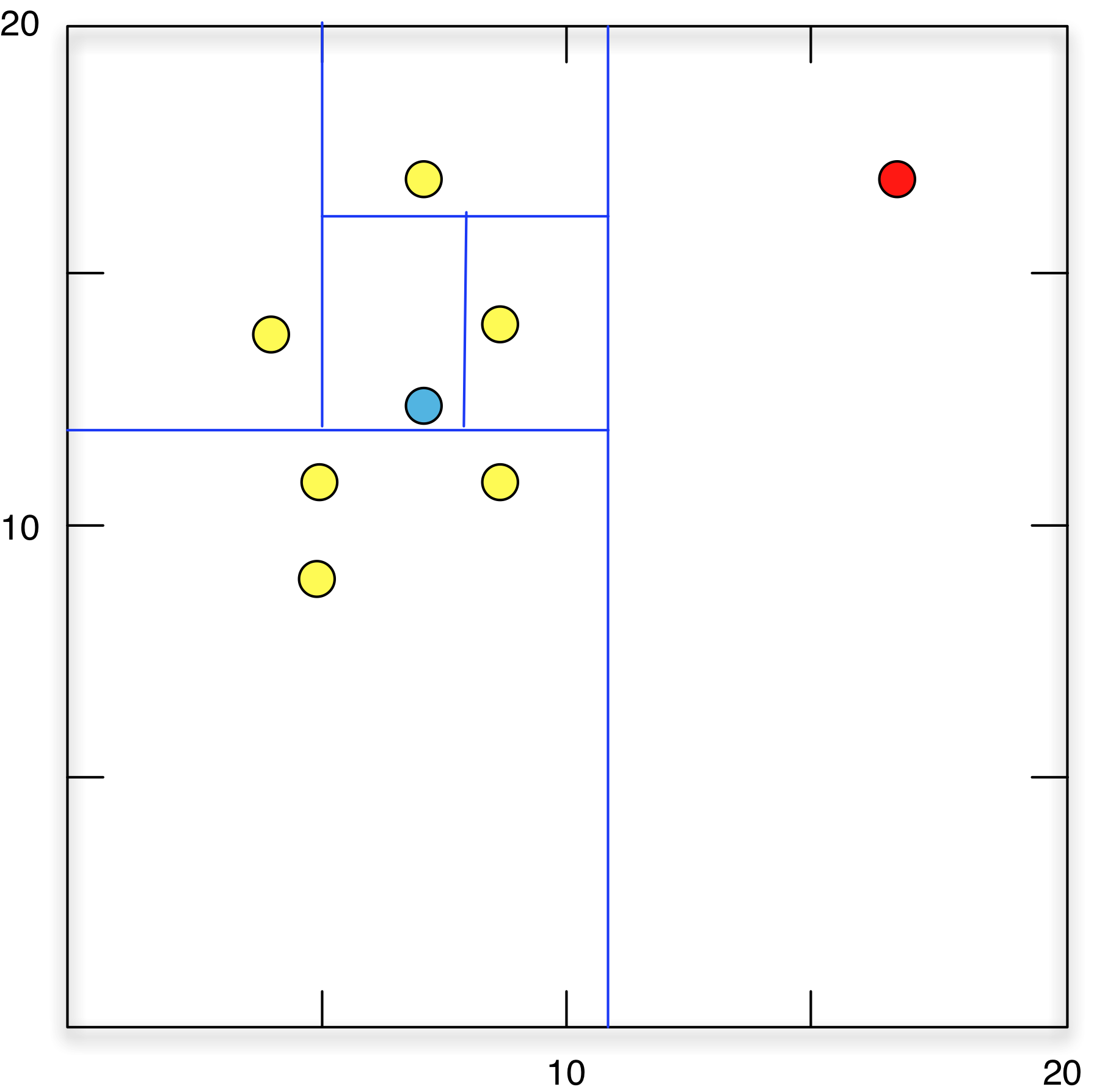}}
\subfigure[]{\includegraphics[width=0.45\textwidth]{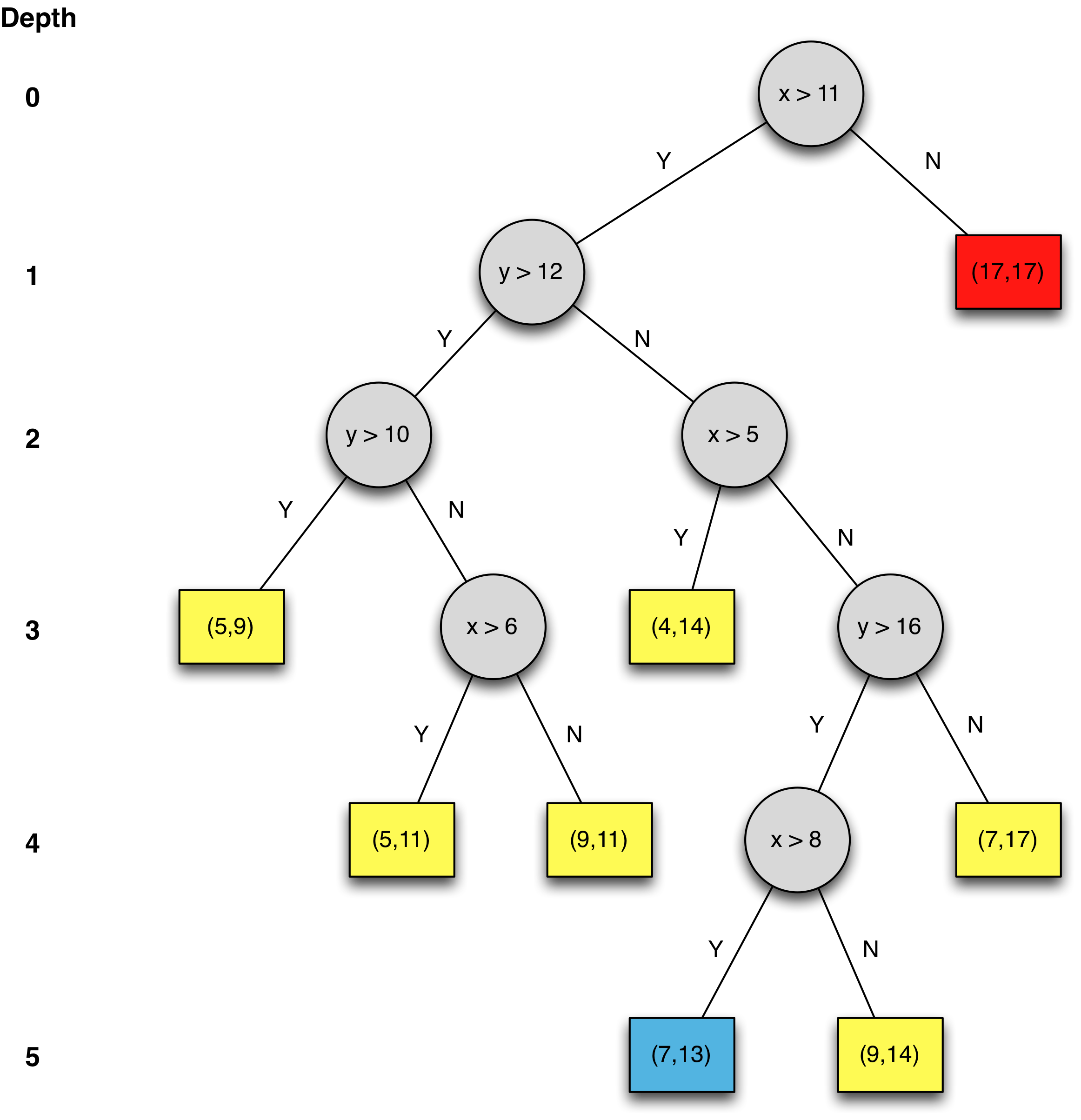}}

\caption{(a) In this example dataset, a randomly constructed binary search
tree isolates the anomalous value (17,17) (shown in red) in just one division,
whereas the medoid value (7,13) (shown in blue) is isolated in five random divisions.
(b) The anomaly has a tree depth of 1, compared to 5 for the medoid point.}
\label{fig:iforestexample}
\end{figure}

Isolation forest  (iForest) \cite{liu2008isolation} is an anomaly detection algorithm.
The algorithm utilises the observation that if a dataset is organised
into a binary search tree, anomalies are more likely to be inserted
at a lesser depth in a tree, compared to non-anomalous values (see Figure~\ref{fig:iforestexample}).
For a given dataset, the algorithm takes $n$ random samples of size $m$.
For each random sample, a binary search tree is constructed, randomly selecting
a dimension and partition point for each comparison node in the tree.
To calculate the anomaly score of a new data point, it is inserted
into each of the $n$ random trees. The anomaly score is derived
from the mean insertion depth across all of the trees.

Isolation forest has a number of advantages as an anomaly detection algorithm:
\begin{itemize}
\item
It requires relatively small samples from large datasets to derive an anomaly detection
function. This makes it fast and scalable.
\item
It does not require example anomalies in the training data set.
\item
Its distance threshold for determining anomalies is based on tree depth,
which is independent of the scaling of the data set dimensions.
\end{itemize}

\subsection{Extended Isolation Forest to categorical data}
\label{if_extend}

Isolation forest was originally proposed for datasets with continuous dimensions.
In this paper we extend the algorithm to consider categorical data.
Our method only requires that for each categorical dimension, values
have an ordering. The ordering may be arbitrary. Each value is then
mapped to a numeric value, based on its ordering.

For example the values \emph{true} and \emph{false} may be mapped
to $\mathit{false} = 0$, $\mathit{true} = 1$.

Having mapped the categorical values to numeric values, the
categorical dimensions can be treated the same way as the
numeric dimensions in the iForest algorithm.

\subsection{Features Extraction}
\label{sec:feature}
The original format of the log files has a total of 42 fields. Some fields are missing in the dataset, such as location information. In this paper, we explore five fields. They are listed below.

\subsubsection{Match Rule}
\begin{figure}[ht]
    \centering
    \includegraphics[height=5cm]{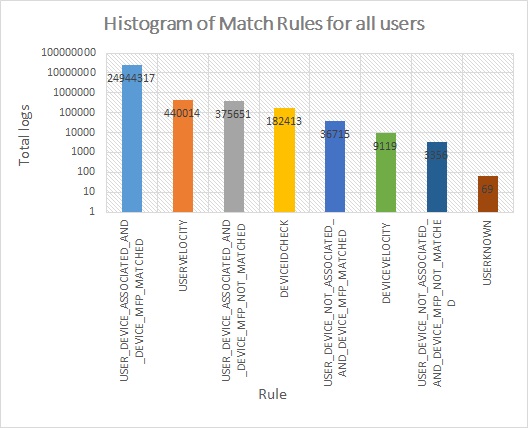}
    \caption{The histogram of Match Rules}
    \label{fig:match}
\end{figure}

The match rule is the rule used in the system for match authentication. 8 predefined rules for all logs are listed below and their distribution is shown in Figure~\ref{fig:match}.
{\tt \small
\begin{itemize}
		\item DEVICEIDCHECK
		\item DEVICEVELOCITY
		\item USER\_DEVICE\_ASSOCIATED\_AND\_DEVICE\_MFP\_MATCHED		
		\item USER\_DEVICE\_ASSOCIATED\_AND\_DEVICE\_MFP\_NOT\_MATCHED
		\item USER\_DEVICE\_NOT\_ASSOCIATED\_AND\_DEVICE\_MFP\_MATCHED
    \item USER\_DEVICE\_NOT\_ASSOCIATED\_AND\_DEVICE\_MFP\_NOT\_MATCHED		
		\item USERVELOCITY
		\item USERKNOWN
	
\end{itemize}
}

\subsubsection{Signature Check}
This field is the information on whether the device signature matched with the incoming device ID. If it matches, the status is ``Y''. Otherwise, it is ``N''. If a Device Check's value is YN or YY, this field can only be Y.
 
\subsubsection{Device Check}
This field checks whether the device is associated with the user. There are three values, NN, YN and YY. NN means that there is no device id on the device.  Either we are seeing this device for the first time or the user deleted cookies. YN means that the device id has been read but it is not associated with the user. It is a known device and a new user, either because this is a shared device or because the user has tried access before but did not successfully authenticate themselves and complete the transaction.
YY means device is associated with the user. 

\subsubsection{Browser}
\begin{figure}[ht]
    \centering
    \includegraphics[height=5cm]{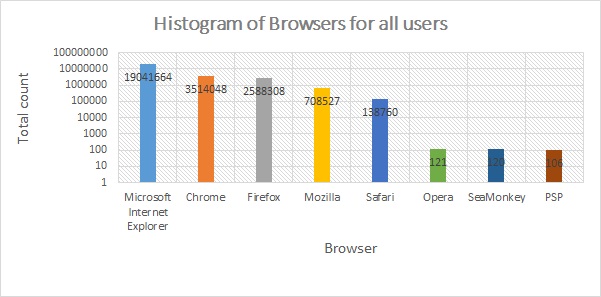}
    \caption{The histogram of Browsers}
    \label{fig:browser}
\end{figure}

This is the browser on which the user logged. This information does not exist as a separate field in the original dataset but can be extracted from the Device Signature field in the log during the pre-process period. There are 8 browsers have been detected. The details of them are listed in Figure~\ref{fig:browser}. The most popular browser is Microsoft Internet Explorer that has more than 19 million logs, followed by Chrome that has more than 3.5 million logs. The three least popular browsers are Opera, SeaMonkey and PSP. Each of them has just over 100 logs.

\subsubsection{Log Time}
This is the Date/Time in GMT that the transaction was evaluated. The original format of timestamps is ``mm/dd/yyyy H:M:S'' or ``mm/dd/yy H:M:S''. We extract two time information from it in this research. They are date of the week and hour of the day.

\section{Experimental Results}
\label{sec:experiment}
Experiments have been carried out to compare the performances of system using both single feature and combined features. For evaluation purpose in this research, for a specific user, we assume all logs of this user are normal and all logs of other users are anomalous for this user. Using this classification, we measure the detection effectiveness. In the real system, once the user model has been built, any testing log with an anomaly score over the threshold will be reported as anomalous. 

\subsection{Data Sets}
All experiments have been carried out on a real enterprise dataset, CA RiskMinder user payroll access logs. This dataset contains a total of 25,991,654 logs, covering a total of 89 days over 14 weeks starting from 15/02/2014. 

\begin{figure}[ht]
    \centering
    \includegraphics[height=5.5cm]{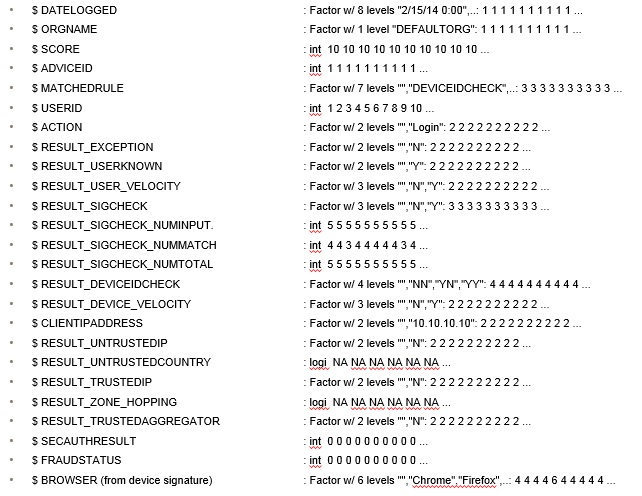}
    \caption{The sample fields of the dataset}
    \label{fig:fields}
\end{figure}

Each log in the file contains 42 fields. Some sample fields with possible values are listed in Figure~\ref{fig:fields} 

\begin{figure}[ht]
    \centering
    \includegraphics[height=5cm]{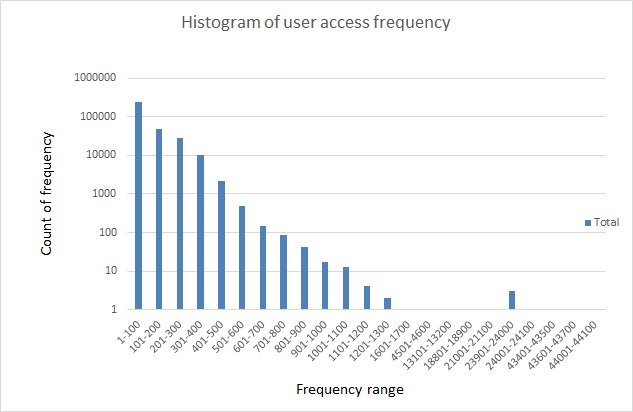}
    \caption{The histogram of user access frequency}
    \label{fig:user_distribution}
\end{figure}

These logs are from 332,672 users. The histogram of user access frequency is shown in Figure~\ref{fig:user_distribution}. As we can seen, except a small amount of users accessed the system very frequently with over 10,000 accesses, most users only accessed a few hundreds times to the system. This research select all users that have access frequency between 501 and 600 and build a user model for each of these users. There are total of 495 users with 267,762 logs in the selected range. We select this range because there is a reasonable amount of users in this range and these logs contain information that is sufficient for testing. We assume that most high frequently access users might be bots and most low frequently access users don't provide enough information. During the testing stage, we randomly select logs from the whole dataset and calculate their anomaly scores against each user model. 

\subsection{Evaluation Metrics}
\label{sec:class-metrics}
When comparing the performance of different classification techniques, it is important to assess how well a classification model is able to correctly assign records to the actual classes. Several metrics are conventionally in use to numerically quantify the effectiveness of classification performance.

To introduce the metrics, let us define that for a class $y_j$, a record is \emph{positive} if it is predicted to belong this specific class and is \emph{negative} if it is predicted to belong other classes. Suppose that for a test set with $n$ records, the set of positive records and negative records for the class are known (for example, as the result of human judgment), and $P$ and $N$ are the number of positive records and negative records respectively, $n = P + N$. Using four important counts~\cite{tk08:pr} defined below, $P = TP + FN$ and $N = FP + TN$.

\begin{itemize}
	\item \emph{TP} represents the true positives which is the number of positive records correctly identified as specific class.
	\item \emph{FP} represents the false positives, the number of negative records which do not belong to the class but were incorrectly identified as it.
	\item \emph{TN} represents the true negatives which refers the number of negative records correctly identified as other classes.
	\item \emph{FN} represents the false negatives, that is the number of positive records which belong to the class but were incorrectly identified as other classes.
\end{itemize}

\subsubsection{Confusion Matrix}
The \textit{confusion matrix} provides information needed to determine how well a classification model performs. In Table~\ref{tab:confusion}, classification results of two classes, A and B, are listed in a 2x2 confusion matrix. For the testing set of total $n = a+b+c+d$ records, class A includes $P = a + b$ positive records and $N = c + d$ negative records while
\[
TP = a \text{ ; } FP = c \text{ ; } TN = d \text{ ; and } FN = b. 
\]

Similarly, class B contains $P = c + d$ positive records and $N = a + b$ negative records while
\[
TP = d \text{ ; } FP = b \text{ ; } TN = a \text{ ; and } FN = c.
\]

\begin{table}
    \begin{center}
    \begin{tabular}{cc|cc}
    	\hline
      & & \multicolumn{2}{c}{Predicted Class} \\ 
      & Class & \texttt{A} & \texttt{B}\\
      \hline
      \hline
      \multirow{2}{15mm}{\parbox{15mm}{Actual Class}} & \parbox{5mm}{\texttt{A}} & $a$ & $b$\\ 
      & \parbox{5mm}{\texttt{B}} & $c$ & $d$\\
      \hline 
    \end{tabular}
    \caption[Confusion matrix for a binary classification problem]{Confusion matrix for for a binary classification problem}  
    \label{tab:confusion}
  \end{center}
\end{table}

\subsubsection{One-dimensional Measures}
In addition to the confusion matrix, some single-figure measures of effectiveness have been developed. In many cases, these single measures are more attractive because of their compactness. In this paper, we used five single measures, namely \emph{true positive rate} (TPrate), \emph{false positive rate} (FPrate), \emph{precision},\emph{recall} and \emph{Accuracy} . They are calculated~\cite{tk08:pr} as:
\begin{equation}
	\begin{array}{l}
		TPrate = \displaystyle \frac{TP}{P} = \displaystyle \frac{TP}{TP + FN} \\
		\\
		FPrate = \displaystyle \frac{FP}{N} = \displaystyle \frac{FP}{FP + TN} \\
	\end{array}
\end {equation}
and
\begin{equation}
	\begin{array}{lcl}
		Precision &=& \displaystyle \frac{TP}{TP + FP} \\
		&&\\
		Recall &=& \displaystyle \frac{TP}{TP + FN} \\
		&&\\
		Accuracy &=& \displaystyle \frac{TP + TN}{TP + FN + FP + TN} \\
	\end{array}
\end {equation}

\subsection{Comparison of Results}
Experiments were carried out to evaluate our approach. In the dataset, there are over 26 million logs from over 330 thousands users. As we discussed before, we select users who have 501-600 access logs. In total, there are 495 users. 

\subsubsection{Comparison of systems on classifying users}
\label{sec:classifyUser}
In this study, we compare 7 detection systems. All systems implement the extended Isolation Forest Algorithm but use different feature sets. There are 5 basic features are selected for the study, namely Match Rule, Signature Check, Device Check, Browser and Log Time. The detail of these features are discussed in Section~\ref{sec:feature}. In summary, System 1-5 use a single basic feature,  System 6 and 7 are based on combined feature set: System 6 combines the first four basic features; System 7 combines all basic features. 

For each system, it runs 10 times on all selected 495 users. In each run, for each user, all records are randomly partitioned into two sets, 100 records as the test set and the rest as the training set. The training set is used to build the Isolation Forest tree as the baseline user model. During the testing period, another 100 records are randomly selected from the whole dataset except this user's. So the test set has a total of 200 records and consists two parts, 100 records classified as this user and 100 records classified as the others. All testing logs are mapped into the Isolation Forest tree and then a anomaly score will be calculated for each one. If the score is high than the threshold, this record is anomaly. The threshold in this research is set at 0.80. If the anomalous record belongs to this user, it is regarded as a False Negative. Otherwise, it is a True Negative. 

In each run, the mean value of 495 users' effectiveness is calculated and the overall performance of each system is calculated as the average of 10 random runs. The summary of the different systems' performance is shown in Table~\ref{tab:detection_results}.
\begin{table}
    \begin{center}
    \begin{tabular}{l c c c c c}
	    	\hline
	    	Packer & TP rate & FP rate & Precision & Recall & Accuracy\\
				\hline
				\hline
				
				{System 1: Single feature Match Rule} & $99.08\%$ & $98.38\%$ & $50.18\%$ & $99.08\%$ & $50.35\%$ \\     
				{System 2: Single feature Signature Check} & $99.78\%$ & $99.08\%$ & $50.18\%$ & $99.78\%$ & $50.35\%$ \\      
				{System 3: Single feature Device Check} & $99.91\%$ & $99.75\%$ & $50.04\%$ & $99.91\%$ & $50.08\%$ \\   
				{System 4: Single feature Browser} & $99.80\%$ & $94.03\%$ & $52.11\%$ & $99.80\%$ & $52.88\%$ \\    
				{System 5: Single feature Time} & $99.46\%$ & $98.54\%$ & $50.24\%$ & $99.46\%$ & $50.46\%$ \\
				{System 6: Combined 4 features} & $99.02\%$ & $94.85\%$ & $51.43\%$ & $99.02\%$ & $52.08\%$ \\
				{System 7: Combined 4 features + time} & $98.92\%$ & $97.50\%$ & $50.50\%$ & $98.92\%$ & $50.77\%$ \\      
				
				\hline
				\end{tabular}
    \caption{Detailed detection results using single feature or combined features}  
    \label{tab:detection_results}
  \end{center}
\end{table}	

Though the combined feature system has slightly better performance than the single feature system, the results show that all tested systems could not distinguish behaviors carried by different users. This is because most features are extracted from categorical data that has very small variants. As presented in Figure \ref{fig:match} and \ref{fig:browser}, the counts of some feature values of match rules and browsers are extreme high that means most users have similar feature sets. It is hard to use Isolation Forest algorithm to isolate anomaly values. 

\subsubsection{Detecting anomalous user behavior}

\begin{figure}[ht]
    \centering
    \includegraphics[height=5cm]{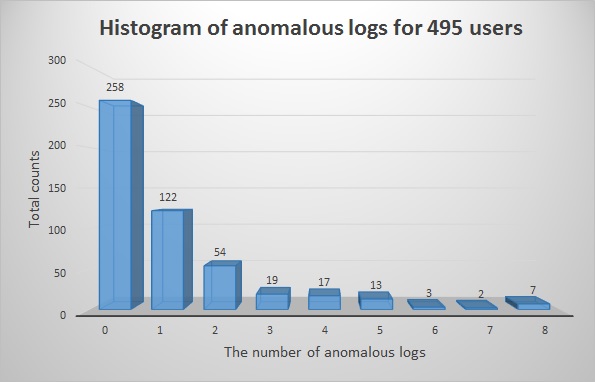}
    \caption{The histogram of false positive value of 495 users using a combined 4 features detection system. A false positive indicates an anomalous behavior of this user.}
    \label{fig:user_fp}
\end{figure}

It is noted that there are logs with high anomaly score detected from both user and others. So the details of each run are examined more closely. Figure~\ref{fig:user_fp} gives the histogram of the number of anomalous logs of all 495 users within one run using the system 6 described as above. 258 out of 495 users don't have any anomalous behavior, followed by 122 users have only one anomalous behavior, such as change the browser or the match rule.

\begin{figure}[ht]
    \centering
    \includegraphics[height=5cm]{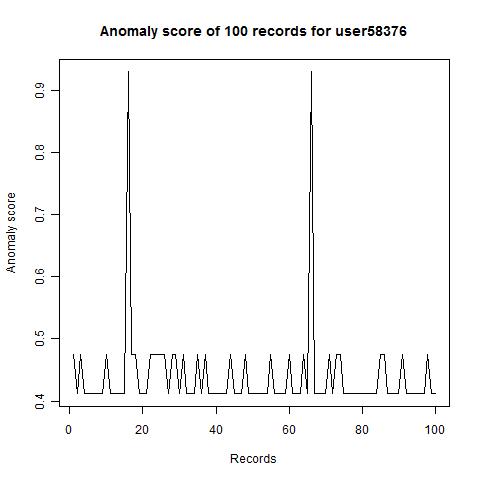}
    \caption{Anomaly scores of 100 records for user58376. It is noted that there are two false positive instances.}
    \label{fig:anomaly}
\end{figure}

An output of one user run is shown in Figure~\ref{fig:anomaly}. As we can see, there are two anomalous records for user 58376 in this run. An example feature sets of this user's normal and abnormal behavior are listed in Table \ref{tab:detection_instances}, In this table, the first two rows report two anomalous behavior.

\begin{table}
    \begin{center}
    \begin{tabular}{r|c|c|c|c|c}
				\hline
	    	Log No & Match Rule & Signature & Device & Browser & Anomaly Score\\
				12303841 & 4 & 1 & 3 & 2 & 0.9307 \\
				4745560 & 4 & 1 & 3 & 1 & 0.9307 \\
				24592463 & 3 & 2 & 3 & 5 & 0.4742 \\
				2399302 & 3 & 2 & 3 & 1 & 0.4118 \\
				18966962 & 3 & 2 & 3 & 5 & 0.4742 \\
				9829951 & 3 & 2 & 3 & 1 & 0.4118 \\
				11434417 & 3 & 2 & 3 & 1 & 0.4118 \\
        \hline
				\end{tabular}
    \caption{Two anomalous behavior and five normal behavior for user 58376}  
    \label{tab:detection_instances}
  \end{center}
\end{table}

				%

\section{Conclusions}
\label{sec:conclusion} 

We have demonstrated that isolation forest is an effective algorithm
for detecting anomalous user behaviour. Our method does not
require any example anomalies in the training data. In this paper we propose a simple
method for extending the isolation forest algorithm to data sets which include categorical
dimensions. Our application of our method to an enterprise dataset
showed promising results. Our experiments tested whether we could
accurately distinguish between the behaviour patterns of
different users. Using five principal features from our
dataset we obtained a recall of approximately 98.92\%. We are able to detect anomalous behavior carried out by the user. We noted that access time is a significant feature, as
users have different patterns of when they usually access the system.
In this work we considered each access as an individual event. Future
work will try to improve the overall accuracy by considering sequences
of events.

\bibliographystyle{unsrt}
\bibliography{av,iforest}

\end{document}